\documentclass[11pt]{article}

\usepackage[margin=1.1in]{geometry}
\usepackage{amsmath,amssymb}
\usepackage{graphicx}
\usepackage[numbers]{natbib}
\usepackage[colorlinks=true,linkcolor=blue,citecolor=blue,urlcolor=blue]{hyperref}

\title{Joint Optimization of Human Headcount and Stochastic AI Resource Capacity}

\author{Marco Montes de Oca\\EnFi, Inc.\\
\texttt{marco.montesdeoca@enfi.ai}}
\date{}

\begin{document}

\maketitle

\begin{abstract}
Generative AI tools are rapidly becoming standard for coding and other business operations. However, while these tools
can drive productivity, their unpredictable, and often significant, costs are putting severe financial strain on
organizations. The leadership of such organizations must
split a fixed budget between
human headcount and generative-AI token capacity, yet standard deterministic
planning ignores both the heavy-tailed volatility of token consumption and the
cognitive cost of auditing AI-generated output. We model the joint allocation
by intersecting a cognitive-friction-adjusted production function with a
chance-constrained stochastic budget frontier. Per-engineer token usage is
treated as i.i.d., non-negative, and right-skewed; no parametric family is
assumed, as only its mean, variance, and skewness enter the analysis. The
chance constraint is reduced to a deterministic equivalent via the Central
Limit Theorem with a Cornish--Fisher skewness correction. Solving the resulting Lagrangian yields a closed-form
optimal headcount, a transcendental condition for optimal per-capita token
intensity, and a bordered-Hessian second-order condition that holds for any
reasonable volatility-to-headcount ratio. Comparative statics show that rising
individual token volatility raises optimal per-capita token intensity while
contracting headcount: although human labor and tokens are complements in
production, the stochastic budget makes them substitutes at the margin.
Individual usage skewness acts as a pure deadweight tax that shrinks the budget
without altering the substitution dynamics. Both conclusions presuppose that
usage dispersion is invariant to the planned mean allocation. In the case where dispersion
instead scales proportionally with the mean, the substitution reverses, so
which regime applies is a sharp, empirically testable question.
\end{abstract}

\begin{sloppypar}
\end{sloppypar}

\begin{figure}[t]
  \centering
  \includegraphics[width=\textwidth]{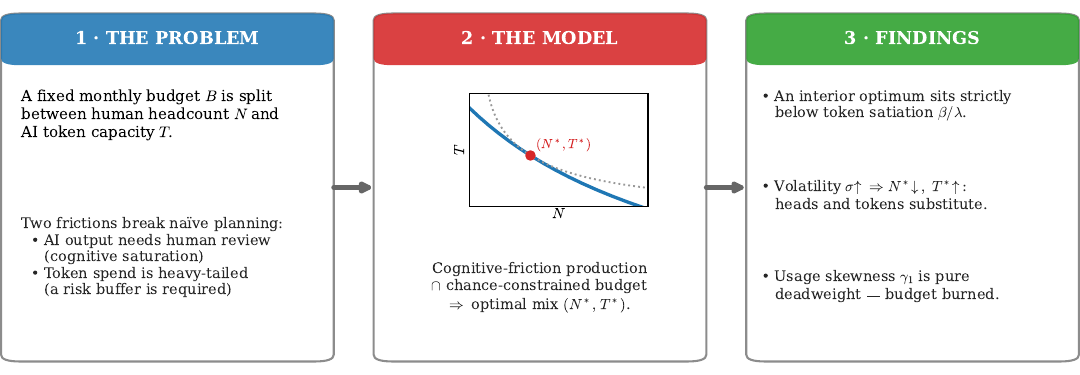}
  \caption{A fixed monthly budget is split between human
  headcount $N$ and AI token capacity $T$ under two frictions: AI output
  must be human-reviewed (cognitive saturation) and per-engineer token spend
  is heavy-tailed (a risk buffer is required). Intersecting a
  cognitive-friction production function with a chance-constrained budget
  yields an interior optimum $(N^*,T^*)$. The two headline comparative
  statics follow: rising usage volatility substitutes headcount for
  per-capita token intensity, and usage skewness is a pure budget deadweight.}
  \label{fig:teaser}
\end{figure}

\section{Frame of Reference and Core Assumptions}

When scaling modern software engineering
organizations\footnote{We focus on software creation, but the same dynamics apply
to other business functions that may benefit from
generative AI tool use.}, leadership must
allocate a fixed monthly budget ($B$) across two primary variable
inputs: human labor (headcount, $N$) and generative AI computational
resources (average monthly tokens per capita, $T$).
In collective-intelligence terms, this is a question of optimal
\emph{composition}: how many human minds should the collective contain,
and how much machine throughput should flow through each~\cite{woolley2010,
bansal2021}.
Traditional deterministic financial planning treats both inputs as
predictable line items, but this assumption
breaks down due to two structural characteristics:

\begin{itemize}
\item
  \textbf{Stochastic Token Volatility:} Individual token
  consumption is highly volatile and prone to extreme right-hand tail
  events (e.g., recursive agent loops, automated mass-refactoring
  scripts, or deep repository re-indexing).
\item
  \textbf{Cognitive Saturation:} Generative AI is not a frictionless
  substitute for human labor. AI engines generate raw code drafts,
  context fragments, and configurations that demand human cognitive
  cycles to audit, debug, and validate. Large token volume
  shifts a human's role from an active builder to a bottlenecked reviewer. This is the classic ``irony of
  automation''~\cite{bainbridge1983}, in which automating most of a task leaves
  the human holding its hardest residual. Our production function
  below captures this phenomenon quantitatively.
\end{itemize}

We model the problem of deriving the optimal balance between human labor and token
consumption as a stochastic program where the goal is to maximize a \textbf{Cognitive-Friction Adjusted Production
Function} subject to a \textbf{Chance-Constrained Stochastic Budget Constraint}.
Figure~\ref{fig:teaser} summarizes this setup and the three findings the
analysis delivers.

\subsection{Core Modeling Assumptions}

\begin{enumerate}
\item
  \textbf{The i.i.d. Resource Pool:} Individual engineer token
  consumption profiles ($T_i$) are independent and identically
  distributed (i.i.d.) non-negative random variables with finite mean
  $\mu$, variance \(\sigma^2\), and strictly positive standardized
  skewness \(\gamma_1\), encoding high-token usage events through a long right-hand tail.
  No parametric family is assumed: the derivations below consume
  only these first three moments, so any standard non-negative
  right-skewed distribution, such as the log-normal or the gamma,
  serves equally well as the motivating shape.
\item
  \textbf{Asymptotic Convergence Qualification:} The deterministic
  equivalent mapped via the Central Limit Theorem paired with the
  Cornish--Fisher expansion is an asymptotic approximation. For small
  teams (small \(N\)), where individual skewness is most
  pronounced and the risk buffer is largest, the convergence rate is
  slower, meaning the boundary acts as an approximation that tightens as
  \(N\) scales.
\item
  \textbf{Exogenous Volatility:} The dispersion parameters \(\sigma\)
  and \(\gamma_1\) are exogenous: invariant to the chosen operational
  mean \(T\). Note that this assumes nothing about volatility being
  controlled or limited, only
  that it does not vary with the planner's allocation, as would be
  plausible if dispersion is driven by infrastructure-level events
  (runaway agent loops, mass refactors) rather than by planned
  intensity. This is a substantive scope condition rather than a mere
  analytical convenience: under a fixed-shape family (constant
  coefficient of variation), \(\sigma\) would instead scale with the
  mean and the marginal substitution results would reverse in sign, as
  Section~6.2 makes explicit.
\item
  \textbf{Strict Binding Constraint:} Because the marginal productivity
  of human headcount is strictly positive across the entire operational
  domain, the organization will always fully exhaust its budget space at
  equilibrium to maximize output.
\end{enumerate}

\section{The Production Field and Cognitive Friction}

Let total organizational software output \(Q\) be mapped as a smooth
scalar field over human labor and total organizational token usage
\(\mathcal{T}\), where \(Q: \mathbb{R}^2_{++} \to \mathbb{R}_+\).
To capture true production-side coupling where individual cognitive
overhead is governed by the per-capita density of the computational
resource mix, we model the system using a Cobb--Douglas core~\cite{cobb1928}
augmented with an exponential friction term:

\begin{equation}Q(N, \mathcal{T}) = A N^\alpha \mathcal{T}^\beta e^{-\lambda \left(\frac{\mathcal{T}}{N}\right)}\end{equation}
Where:

\begin{itemize}
\item
  \(A > 0\): The baseline structural efficiency of the engineering
  environment (codebase architecture cleanliness, documentation
  standards, automated testing maturity).
\item
  \(0 < \alpha, \beta < 1\): The output elasticities of human labor
  and token leverage, respectively, establishing standard diminishing
  marginal returns for both inputs.
\item
  \(\lambda > 0\): The \textbf{Cognitive Overload Coefficient}
  (dimensioned in \(\text{tokens}^{-1}\)), representing the exact rate
  at which human oversight capacity is saturated by processing
  AI-generated context.
\end{itemize}

To evaluate this from an operational management perspective, we
transition to the per-capita token convention by substituting
\(\mathcal{T} = N \cdot T\), where \(T\) is the average monthly token
consumption per engineer:

\begin{equation}\label{eq:production}Q(N, T) = A N^\alpha (N T)^\beta e^{-\lambda \left(\frac{N T}{N}\right)} = A N^{\alpha + \beta} T^\beta e^{-\lambda T}\end{equation}

\subsection{Step-by-Step Derivation of Marginal Productivities}

We take the partial derivatives of \(Q(N, T)\) with respect to
our decision coordinates.

\paragraph{Marginal Productivity of Tokens:}


\begin{equation}\frac{\partial Q}{\partial T} = A N^{\alpha + \beta} \left( \frac{\partial}{\partial T}[T^\beta] \cdot e^{-\lambda T} + T^\beta \cdot \frac{\partial}{\partial T}[e^{-\lambda T}] \right)\end{equation}

\begin{equation}\frac{\partial Q}{\partial T} = A N^{\alpha + \beta} \left( \beta T^{\beta-1} e^{-\lambda T} + T^\beta (-\lambda e^{-\lambda T}) \right)\end{equation}

\begin{equation}\frac{\partial Q}{\partial T} = A N^{\alpha + \beta} T^\beta e^{-\lambda T} \left( \frac{\beta}{T} - \lambda \right)\end{equation}
Substituting the original production function~\eqref{eq:production} back into the
equation isolates the final gradient:

\begin{equation}\frac{\partial Q}{\partial T} = Q(N, T) \cdot \left[ \frac{\beta}{T} - \lambda \right]\end{equation}

\paragraph{Marginal Productivity of Labor:}

Differentiating directly with respect to the headcount term \(N\),
treating the per-capita token allocation \(T\) as a fixed operational
intensity:

\begin{equation}\frac{\partial Q}{\partial N} = (\alpha + \beta) A N^{\alpha + \beta - 1} T^\beta e^{-\lambda T}\end{equation}
Multiplying the numerator and denominator by \(N\) reveals its
proportional relationship to total output:

\begin{equation}\frac{\partial Q}{\partial N} = \frac{(\alpha + \beta) A N^{\alpha + \beta} T^\beta e^{-\lambda T}}{N} = Q(N, T) \cdot \left[ \frac{\alpha + \beta}{N} \right]\end{equation}

\subsection{Production Complementarity and Satiation}

While the production function becomes multiplicatively separable in
\(N\) and \(T\) after substitution, the inputs remain tightly coupled
via their cross-partial derivative:

\begin{equation}\frac{\partial^2 Q}{\partial N \partial T} = \frac{\partial}{\partial N}\left[ Q(N, T) \cdot \left(\frac{\beta}{T} - \lambda\right) \right] = \frac{Q(N, T) \cdot (\alpha + \beta)}{N}\left[\frac{\beta}{T} - \lambda\right]\end{equation}
Below the physical satiation threshold (\(T < \beta/\lambda\)), the
cross-partial derivative is strictly positive
(\(\frac{\partial^2 Q}{\partial N \partial T} > 0\)). This
establishes \(N\) and \(T\) as structural \textbf{Edgeworth complements
in production}~\cite{samuelson1947,milgrom1990}. This means that hiring more human engineers directly expands the
marginal productivity of tokens, as a larger human pool provides
more cognitive capacity to safely ingest and ship AI output.

Setting the token gradient \(\frac{\partial Q}{\partial T} = 0\)
isolates the unconstrained token satiation point:

\begin{equation}\label{eq:tsat}T_{\text{sat}} = \frac{\beta}{\lambda}\end{equation}
This limit marks the structural peak where further token consumption
yields a net decrease in organizational output due to cognitive
overload. Because tokens carry a real financial cost (\(P_t > 0\)),
the budget-constrained economic optimum (\(T^*\)) will sit strictly
below this physical ceiling~\eqref{eq:tsat} (\(T^* < \beta/\lambda\)).
\textbf{Intuitive Example:} Imagine an engineering organization where
\(\beta = 0.4\) and \(\lambda = 8 \times 10^{-9}\,\text{tokens}^{-1}\)
(equivalently \(0.008\,\text{MTok}^{-1}\), the unit used in the figures). The
unconstrained satiation point sits at
\(T_{\text{sat}} = 0.4 / (8 \times 10^{-9}) = 5 \times 10^{7}\) or
roughly 50 million tokens/engineer/month, which is well within the range of present-day
agentic usage. If tokens were entirely free (\(P_t = 0\)), leadership should
cap individual allocation at that ceiling. Beyond it, engineers spend so much
time reviewing automated AI output that total software shipping velocity
slows down.

\begin{figure}[t]
  \centering
  \includegraphics[width=0.78\linewidth]{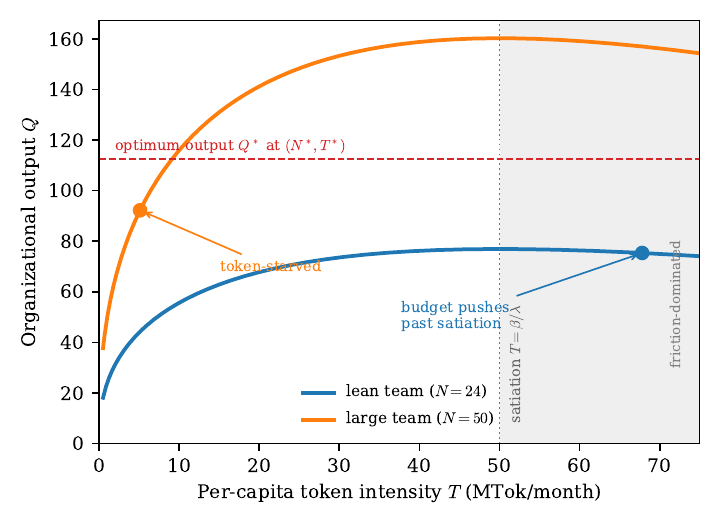}
  \caption{Output versus token intensity at fixed headcount, for a lean
  ($N=24$) and a large ($N=50$) team. Each curve is the production function
  $Q(N,T)=AN^{\alpha+\beta}T^\beta e^{-\lambda T}$; the exponential
  cognitive-friction factor makes output rise, peak at the satiation intensity
  $T=\beta/\lambda$ (dotted line), and decline in the friction-dominated region
  (shaded). Dots mark each team's budget-feasible operating point at the same
  budget: the lean team can afford so many tokens per engineer that it is driven
  \emph{past} satiation, while the large team is \emph{token-starved} far to the
  left. Neither reaches the output $Q^*$ of
  the balanced optimum $(N^*,T^*)$ (dashed). Same illustrative parameters as
  Fig.~\ref{fig:tension}.}
  \label{fig:satiation}
\end{figure}

Figure~\ref{fig:satiation} makes this satiation concrete and previews the
central substitution: at a fixed budget, a leaner team is driven \emph{past}
the satiation peak into the friction-dominated regime, while a larger team is
left token-starved, so neither matches the balanced optimum of Section~5.

\section{Derivation of the Stochastic Budget Frontier}

The organization operates under a hard monthly budget \(B\). Let \(S\)
represent the average individual engineer salary and \(P_t\) represent
the unit price per token.
Because individual monthly token consumption values (\(T_i\)) are
random variables, total expenditure is also a random variable. To
protect the organization against cost overruns, leadership establishes
an acceptable risk tolerance threshold \(\epsilon\) (e.g.,
\(\epsilon = 0.05\), representing a strict 5\% maximum allowable risk
that token consumption spikes will cause a budget breach in any given
month). This yields the following chance constraint~\cite{charnes1959}:

\begin{equation}P\left(N \cdot S + P_t \sum_{i=1}^{N} T_i \le B\right) \ge 1 - \epsilon\end{equation}

\subsection{Step-by-Step Transformation to a Deterministic Equivalent}

To make this probabilistic statement useful for algebraic optimization,
we transform it into a deterministic equivalent using standard
statistical properties.

\paragraph{Step 1: Isolate the Random Sum}

We isolate the summation of individual token consumption,
\(\sum_{i=1}^{N} T_i\), which contains all the stochastic volatility
in our system. Subtract the fixed human payroll costs (\(N \cdot S\))
and divide by the unit token price (\(P_t\)) to isolate the aggregate
random variable sum:

\begin{equation}P\left(\sum_{i=1}^{N} T_i \le \frac{B - N \cdot S}{P_t}\right) \ge 1 - \epsilon\end{equation}

\paragraph{Step 2: Map Aggregate Moments via the Central Limit Theorem}

According to the Central Limit Theorem (CLT), the sum of independent,
identically distributed random variables approaches a normal
distribution as the sample size \(N\) scales. We calculate the aggregate
expected mean and aggregate variance of our engineering pool:

\begin{itemize}
\item
  \textbf{Aggregate Expected Mean:}
  \(E\left[\sum_{i=1}^{N} T_i\right] = \sum_{i=1}^{N} E[T_i] = N\mu\)
\item
  \textbf{Aggregate Variance:}
  \(\text{Var}\left(\sum_{i=1}^{N} T_i\right) = \sum_{i=1}^{N} \text{Var}(T_i) = N\sigma^2\)
\item
  \textbf{Aggregate Standard Deviation:}
  \(\sqrt{\text{Var}} = \sqrt{N}\sigma\)
\end{itemize}

To maintain mathematical honesty regarding tail behavior in finite
engineering populations, we apply a Cornish--Fisher expansion~\cite{cornish1938} to adjust
for the non-zero population skewness (\(\gamma_1\)). The aggregate
skewness of the team scales as \(\Gamma_1 = \gamma_1 / \sqrt{N}\).
This expansion adjusts the standard normal critical value into a
skewness-adjusted quantile score (\(w_{1-\epsilon}\)):

\begin{equation}w_{1-\epsilon} = z_{1-\epsilon} + \frac{\Gamma_1}{6}(z_{1-\epsilon}^2 - 1) = z_{1-\epsilon} + \frac{\gamma_1}{6\sqrt{N}}(z_{1-\epsilon}^2 - 1)\end{equation}

\paragraph{Step 3: Z-Score Standardization}

We standardize our random sum into a standard normal distribution by
subtracting the aggregate mean (\(N\mu\)) and dividing by the aggregate
standard deviation (\(\sqrt{N}\sigma\)) on both sides of the
inequality:

\begin{equation}P\left( \frac{\sum T_i - N\mu}{\sqrt{N}\sigma} \le \frac{\frac{B - N \cdot S}{P_t} - N\mu}{\sqrt{N}\sigma} \right) \ge 1 - \epsilon \implies \Phi\left( \frac{\frac{B - N \cdot S}{P_t} - N\mu}{\sqrt{N}\sigma} \right) \ge 1 - \epsilon\end{equation}
Applying the inverse quantile function (\(\Phi^{-1}\)) to both sides
isolates the scalar terms against our skewness-adjusted critical value:

\begin{equation}\frac{\frac{B - N \cdot S}{P_t} - N\mu}{\sqrt{N}\sigma} \ge z_{1-\epsilon} + \frac{\gamma_1}{6\sqrt{N}}(z_{1-\epsilon}^2 - 1)\end{equation}

\paragraph{Step 4: Clearing Denominators and Isolating Budget \(B\)}

Multiply both sides of the inequality by the aggregate standard
deviation \(\sqrt{N}\sigma\):

\begin{equation}\frac{B - N \cdot S}{P_t} - N\mu \ge \left[ z_{1-\epsilon} + \frac{\gamma_1}{6\sqrt{N}}(z_{1-\epsilon}^2 - 1) \right] \sigma \sqrt{N}\end{equation}
Distributing \(\sigma\sqrt{N}\) across the brackets causes the
\(\sqrt{N}\) terms in the skewness modifier to cancel out cleanly:

\begin{equation}\frac{B - N \cdot S}{P_t} - N\mu \ge z_{1-\epsilon}\sigma\sqrt{N} + \frac{\gamma_1 \sigma (z_{1-\epsilon}^2 - 1)}{6}\end{equation}
Multiply the entire equation by the token price \(P_t\), add
\(N \cdot S\) to both sides, and factor out headcount \(N\) from the
baseline cost components. This yields the deterministic-equivalent
budget frontier:

\begin{equation}\label{eq:frontier}B \ge N(S + P_t\mu) + z_{1-\epsilon}P_t\sigma\sqrt{N} + \frac{\gamma_1 P_t \sigma (z_{1-\epsilon}^2 - 1)}{6}\end{equation}

\section{System Optimization and Equilibrium Derivation}

To locate the coordinates that maximize total organizational software
output within our safe financial boundary, we formulate the Lagrangian
function (\(\mathcal{L}\))~\cite{chiang1984}. We define this over our production field
and our binding deterministic budget constraint, treating our
operational per-capita token decision variable \(T\) as the runtime
expected mean \(\mu\):

\begin{equation}\mathcal{L}(N, T, \Lambda) = \left( A N^{\alpha + \beta} T^\beta e^{-\lambda T} \right) + \Lambda \left( B - N(S + P_t T) - z_{1-\epsilon}P_t\sigma\sqrt{N} - K_{\gamma} \right)\end{equation}
Where \(\Lambda\) is the Lagrange multiplier, and
\(K_{\gamma} = \frac{\gamma_1 P_t \sigma (z_{1-\epsilon}^2 - 1)}{6}\)
collects the constant skewness deadweight term.

\subsection{Step 1: Evaluating First-Order Conditions (FOC)}

We take the partial derivatives of the Lagrangian with respect to our
decision variables and set them identically to zero.

\paragraph{Differentiating with respect to Token Intensity (\(\frac{\partial \mathcal{L}}{\partial T} = 0\)):}

\begin{equation}Q(N, T) \cdot \left[ \frac{\beta}{T} - \lambda \right] - \Lambda P_t N = 0 \implies \Lambda = \frac{Q(N, T)}{P_t N} \left[ \frac{\beta}{T} - \lambda \right]\end{equation}

\paragraph{Differentiating with respect to Headcount (\(\frac{\partial \mathcal{L}}{\partial N} = 0\)):}

\begin{equation}Q(N, T) \cdot \left[ \frac{\alpha + \beta}{N} \right] - \Lambda \left( S + P_t T + \frac{z_{1-\epsilon}P_t\sigma}{2\sqrt{N}} \right) = 0 \implies \Lambda = \frac{Q(N, T) \cdot (\alpha + \beta)}{N \left( S + P_t T + \frac{z_{1-\epsilon}P_t\sigma}{2\sqrt{N}} \right)}\end{equation}

\subsection{Step 2: Equating the Marginal Conditions}

Equating our two independent expressions for the marginal value of
capital \(\Lambda\) and canceling out the total output scalar
\(Q(N, T)\) and the headcount parameter \(N\) from both sides yields
the equilibrium condition:

\begin{equation}\label{eq:equilibrium}\frac{\frac{\beta}{T} - \lambda}{P_t} = \frac{\alpha + \beta}{S + P_t T + \frac{z_{1-\epsilon}P_t\sigma}{2\sqrt{N}}}\end{equation}

\section{Closed-Form Solution Space}

To resolve the precise operational headcount (\(N^*\)) and per-capita
token allocation (\(T^*\)), the equilibrium relation must be solved
simultaneously alongside our binding budget boundary as a non-linear
\(2 \times 2\) system. Headcount is treated as continuous throughout;
at the organizational scales considered, rounding \(N^*\) to an integer
perturbs the optimum negligibly.

\begin{figure}[t]
  \centering
  \includegraphics[width=0.72\linewidth]{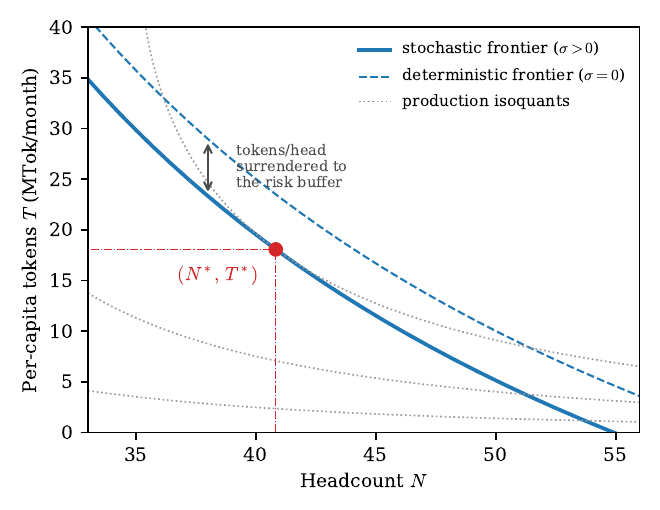}
  \caption{The headcount--token tension. On the binding stochastic budget
  frontier (solid), every additional engineer lowers the affordable
  per-capita token allocation; the optimum $(N^*,T^*)$ is the tangency of
  this frontier with the highest attainable production isoquant. The risk
  buffer holds the stochastic frontier strictly inside the deterministic
  ($\sigma=0$) frontier (dashed). Illustrative parameters (a premium
  agentic-coding regime, tokens in millions): $\alpha=0.6$, $\beta=0.4$,
  $\lambda=0.008\,\text{MTok}^{-1}$ (satiation $\beta/\lambda=50$~MTok),
  $S=\$15$k/month, $P_t=\$300$/MTok, $\sigma=20$~MTok, $\gamma_1=2$,
  $z_{0.95}=1.645$, $B=\$900$k/month.}
  \label{fig:tension}
\end{figure}

Figure~\ref{fig:tension} visualizes this system: the optimum is the point
where the binding budget frontier is tangent to the highest production
isoquant it can reach, and the stochastic risk buffer pulls that frontier
strictly inside its deterministic counterpart.

\subsection{Step 1: Solving for Optimal Headcount (\(N^*\))}

We isolate headcount by rearranging our binding budget boundary~\eqref{eq:frontier} into a
standard quadratic form with respect to the variable \(\sqrt{N}\):

\begin{equation}(S + P_t T)N + (z_{1-\epsilon}P_t\sigma)\sqrt{N} - (B - K_\gamma) = 0\end{equation}
Applying the quadratic formula yields the exact structural mapping for
headcount:

\begin{equation}\label{eq:nstar}\sqrt{N^*} = \frac{-z_{1-\epsilon}P_t\sigma + \sqrt{(z_{1-\epsilon}P_t\sigma)^2 + 4(S + P_t T^*)(B - K_\gamma)}}{2(S + P_t T^*)}\end{equation}

\subsection{Step 2: Solving for Optimal Token Intensity (\(T^*\))}

Cross-multiplying the equilibrium condition~\eqref{eq:equilibrium} from Section~4.2,
canceling out matching \(\beta P_t\) parameters, and grouping
deterministic cost terms on the left side versus risk terms on the right
side yields:

\begin{equation}\frac{\beta S}{T} - \lambda S - \lambda P_t T - P_t \alpha = \frac{z_{1-\epsilon}P_t\sigma}{2\sqrt{N}} \left( \lambda - \frac{\beta}{T} \right)\end{equation}
Multiplying the entire equation by \(-T\) to clear all denominators
leaves the transcendental condition:

\begin{equation}\label{eq:transcendental}\lambda P_t T^2 + \left( \lambda S + P_t \alpha \right) T - \beta S = \frac{z_{1-\epsilon}P_t\sigma}{2\sqrt{N}} \left( \beta - \lambda T \right)\end{equation}

\section{Optimization Rigor \& Comparative Statics}

\subsection{Verification of the Bordered Hessian (Local Extremum)}

To verify that our stationary coordinates \((N^*, T^*)\) represent a
valid local maximum rather than a minimum or saddle point, we evaluate
the determinant of the bordered Hessian matrix \(\mathbf{H}\). With the
binding boundary~\eqref{eq:frontier} written as
\(g(N, T) = N(S + P_t T) + z_{1-\epsilon}P_t\sigma\sqrt{N} + K_{\gamma} - B = 0\),
the relevant second-order partials are:

\begin{equation}\frac{\partial^2 Q}{\partial T^2} = Q \left[\delta^2 - \frac{\beta}{T^2}\right], \quad \frac{\partial^2 Q}{\partial N^2} = \frac{Q(\alpha+\beta)(\alpha+\beta-1)}{N^2}, \quad \frac{\partial^2 Q}{\partial N \partial T} = \frac{(\alpha+\beta)}{N}\frac{\partial Q}{\partial T}\end{equation}

\begin{equation}g_N = S + P_t T + \frac{z_{1-\epsilon}P_t\sigma}{2\sqrt{N}}, \quad g_T = P_t N, \quad g_{NN} = -\frac{z_{1-\epsilon}P_t\sigma}{4 N^{3/2}}, \quad g_{TT} = 0, \quad g_{NT} = P_t\end{equation}
Where we abbreviate the per-token marginal-output ratio
\(\delta \equiv \frac{\beta}{T} - \lambda = \frac{1}{Q}\frac{\partial Q}{\partial T} > 0\)
(strictly positive below satiation). The bordered Hessian is:

\begin{equation}\mathbf{H} = \begin{bmatrix} 0 & g_N & g_T \\ g_N & Q_{NN} - \Lambda g_{NN} & Q_{NT} - \Lambda g_{NT} \\ g_T & Q_{NT} - \Lambda g_{NT} & Q_{TT} \end{bmatrix}\end{equation}

\begin{equation}|\mathbf{H}| = -g_N^2 Q_{TT} + 2 g_N g_T (Q_{NT} - \Lambda P_t) - g_T^2 (Q_{NN} - \Lambda g_{NN})\end{equation}
The decisive simplification eliminates \(g_N\) using the equilibrium
condition~\eqref{eq:equilibrium} (Section~4.2),
\(g_N = \frac{(\alpha+\beta)P_t}{\delta}\). Two intermediate
identities follow directly from the first-order conditions. First, the
cross term is governed by the firm's returns to scale:

\begin{equation}Q_{NT} - \Lambda P_t = \frac{(\alpha+\beta)}{N}\frac{\partial Q}{\partial T} - \frac{1}{N}\frac{\partial Q}{\partial T} = \frac{(\alpha + \beta - 1)}{N}\frac{\partial Q}{\partial T}\end{equation}
Second, the headcount-curvature term carries the budget buffer's
concavity. Substituting these and \(g_T = P_t N\), the
output-proportional components of the determinant telescope cleanly (the
\(m^2\) and \(m(m-1)\) terms collapse, with
\(m \equiv \alpha+\beta\)). Utilizing
\(T^2\delta^2 = (\beta - \lambda T)^2\), the determinant resolves
to a single closed form:

\begin{equation}|\mathbf{H}| = \underbrace{P_t^2 Q (\alpha+\beta) \left[\frac{(\alpha+\beta)\beta}{(\beta - \lambda T)^2} - 1\right]}_{\text{Production Curvature } (>0)} - \underbrace{\frac{P_t^3 \Lambda z_{1-\epsilon}\sigma\sqrt{N}}{4}}_{\text{Budget-Buffer Curvature } (>0)}\end{equation}
This decomposition is exact and assumes nothing about returns to scale.
Two structural facts establish the sign:

\begin{itemize}
\item
  The production-curvature term is unconditionally positive below
  satiation. The factor
  \(\frac{(\alpha+\beta)\beta}{(\beta - \lambda T)^2}\) is
  minimized as \(T \to 0\), where it equals
  \(\frac{\alpha+\beta}{\beta}\); hence the bracket is bounded below
  by \(\frac{\alpha}{\beta} > 0\) and diverges as
  \(T \to \frac{\beta}{\lambda}\). The strong log-concavity of
  output in tokens
  (\(\frac{\partial^2 \ln Q}{\partial T^2} = -\frac{\beta}{T^2} < 0\))
  is the engine driving this term.
\item
  The budget-buffer term enters negatively. Because the risk cushion
  \(z_{1-\epsilon}P_t\sigma\sqrt{N}\) is concave in headcount
  (\(g_{NN} < 0\)), it renders the feasible region locally non-convex
  and acts against the interior maximum. The optimum is stable precisely
  when production curvature dominates this destabilizing force.
\end{itemize}

Equating the two and using \(\Lambda/Q = \delta / (P_t N)\), the
second-order condition \(|\mathbf{H}| > 0\) reduces to a clean,
interpretable inequality:

\begin{equation}(\alpha+\beta)\left[\frac{(\alpha+\beta)\beta}{(\beta - \lambda T^*)^2} - 1\right] > \frac{\left(\frac{\beta}{T^*} - \lambda\right) z_{1-\epsilon}\sigma}{4\sqrt{N^*}}\end{equation}
The left side is order-one and strictly positive; the right side scales
as \(\sigma / \sqrt{N^*}\). The candidate \((N^*, T^*)\) is
therefore a verified local maximum for any reasonable
volatility-to-headcount ratio, failing only in the pathological regime
of extreme individual variance on a very small team (the same weakness noted earlier around our use of the CLT).

For a global statement we reduce to one dimension along the binding
boundary: substituting the closed-form \(N^*(T)\)~\eqref{eq:nstar}
collapses the system to a single continuous objective
\(\varphi(T) = Q(N^*(T), T)\) on the interval
\((0, \beta/\lambda)\). We do not prove \(\varphi\) globally
unimodal; across the illustrative parameter ranges of
Figs.~\ref{fig:satiation}--\ref{fig:variance} it numerically exhibits a
single interior stationary point, which we accordingly treat as the
global solution.

\subsection{Comparative Statics: How Volatility Alters the Mix (\(\frac{\partial T^*}{\partial \sigma}\))}

We analyze the behavior of our transcendental condition~\eqref{eq:transcendental} under
changes in stochastic variance (\(\sigma\)). Let \(T_0\) solve the
deterministic baseline equation where \(f(T_0) = 0\) when
\(\sigma = 0\):

\begin{itemize}
\item
  \(f(T)\) describes an upward-opening parabola with a vertical
  intercept at \(f(0) = -\beta S < 0\). Thus, at its positive root
  \(T_0\), the function is strictly increasing: \(f'(T_0) > 0\).
\item
  \(g(T)\) represents the stochastic modification layer. Because our
  economic optimum must sit below the physical satiation threshold
  (\(T^* < \frac{\beta}{\lambda}\)), the term
  \((\beta - \lambda T)\) is strictly positive across the active
  domain. Hence, for any non-zero variance \(\sigma > 0\),
  \(g(T) > 0\).
\end{itemize}

At the original deterministic baseline, \(f(T_0) = 0 < g(T_0)\).
Because \(f(T)\) is monotonically increasing and \(g(T)\) is
monotonically decreasing on the interval
\(T < \frac{\beta}{\lambda}\), the new intersection point \(T^*\)
must lie strictly to the right of the baseline root:

\begin{equation}T^* > T_0 \implies \frac{\partial T^*}{\partial \sigma} > 0\end{equation}

\begin{figure}[t]
  \centering
  \includegraphics[width=\linewidth]{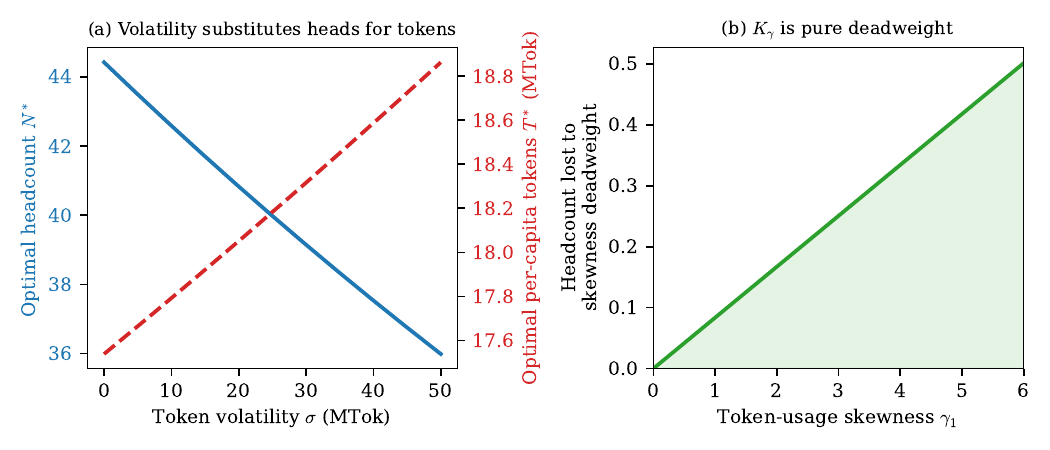}
  \caption{Volatility and skewness erode headcount. \textbf{(a)} As individual
  token volatility $\sigma$ rises, optimal headcount $N^*$ contracts (left
  axis) while optimal per-capita token intensity $T^*$ expands (right
  axis). In this case, the stochastic budget turns two production complements into
  marginal substitutes, so squeezing $\sigma$ is what funds a healthy
  headcount increase. \textbf{(b)} The skewness term $K_\gamma$ is a pure
  deadweight tax: it grows linearly in usage skewness $\gamma_1$, shown here
  as the engineers' worth of budget it consumes, independent of the
  allocation mix. Same illustrative parameters as Fig.~\ref{fig:tension}.}
  \label{fig:variance}
\end{figure}

Two qualifications to keep in mind: First, \(g(T)\) is
evaluated holding \(\sqrt{N}\) fixed, whereas \(N^*(T)\) shifts along
the binding frontier; the local claim, that any \(\sigma > 0\) lifts
\(g\) above \(f(T_0) = 0\) and pushes the root rightward, is
unaffected, and the global monotonicity of \(T^*\) in \(\sigma\) is
confirmed numerically across the range of Fig.~\ref{fig:variance}(a).
Second, the direction of the effect inherits the scope condition of
Assumption~3, to which we return below.

\paragraph{The Substitution Mechanism}

This mathematical validation reveals the core resource allocation
behavior under uncertainty: \textbf{higher individual token volatility
increases your optimal per-capita token intensity while contracting
headcount.} Figure~\ref{fig:variance}(a) traces both optimal coordinates
across the volatility axis, making the substitution explicit: the
healthiest headcount is bought not by spending more but by squashing
$\sigma$.
The active risk term \(\frac{z_{1-\epsilon}P_t\sigma}{2\sqrt{N}}\)
acts as a direct financial tax tied to human headcount. It dictates that
hiring a human engineer requires holding back a defensive cash cushion
to absorb potential token overruns within our risk tolerance
\(\epsilon\). As variance (\(\sigma\)) increases, this safety buffer
expands, making a human hire marginally more expensive.
The optimization engine responds to this risk by substituting away from
now-costlier human headcount (\(N^* \downarrow\)) and driving
remaining resources into higher per-capita token intensity
(\(T^* \uparrow\)). While these inputs are complements in production,
they act as substitutes under a stochastic budget framework. Individual
usage skewness (\(\gamma_1\)), meanwhile, acts as a pure deadweight
tax (\(K_\gamma\)) that reduces absolute budget space uniformly
without altering these marginal substitution dynamics.
Figure~\ref{fig:variance}(b) quantifies this loss directly: every unit of
usage skewness burns a fixed amount of budget (budget that could otherwise fund engineers) regardless of how the remaining
resources are split.

\paragraph{Scope Condition: Where the Volatility Lives}

The direction of this substitution is conditional on Assumption~3:
dispersion, whatever its magnitude, is invariant to the planned mean
allocation. In the opposing fixed-shape regime (a constant
coefficient of variation \(\sigma = cT\), as implied by a log-normal
or gamma family with a frozen shape parameter) the risk buffer becomes
\(z_{1-\epsilon}P_t c T\sqrt{N}\), so volatility taxes token intensity
directly rather than taxing headcount through the \(\sqrt{N}\) buffer.
Re-solving the system numerically in that regime reverses the
substitution: rising \(c\) suppresses \(T^*\) while leaving \(N^*\)
nearly unchanged. The model therefore delivers a sharp, testable
dichotomy rather than one universal direction: whether volatility should
push an organization toward higher or lower token intensity is decided
by whether its spend dispersion scales with the planned allocation or is
independent of it. This is the empirically testable question mentioned in the abstract.
It could, for instance, be measured directly via usage telemetry.

\section{Discussion and Related Work}

Our contribution sits at the intersection of three literatures that, to our
knowledge, have not previously been combined: the economics of production and
input substitution, the empirical study of generative AI as a workplace input,
and chance-constrained stochastic optimization. Treating AI token capacity as a
\emph{stochastic, priced input} jointly optimized with human headcount under a
probabilistic budget is, we believe, novel; the paragraphs below position each
component against its lineage and flag where our assumptions invite scrutiny.

\paragraph{Production, substitution, and complementarity.}
Our output model is a Cobb--Douglas core~\cite{cobb1928} augmented with an
exponential satiation term. Cobb--Douglas is the unit-elasticity special case of
the constant-elasticity-of-substitution (CES) family introduced by Arrow,
Chenery, Minhas, and Solow~\cite{arrow1961}, which holds the elasticity of
substitution between inputs constant; our friction term deliberately breaks that
constancy near satiation. The Edgeworth complementarity we derive between
headcount and tokens (Section 2.2) is the AI-era analogue of \emph{capital--skill
complementarity}, for which Ohanian, Orak, and Shen~\cite{ohanian2023}, extending
the Krusell--Ohanian--R\'ios-Rull--Violante tradition, find continued strong
empirical support: just as equipment capital complements skilled labor, AI token
capacity complements the human reviewers who ship its output.

\paragraph{AI as a productive input, and the review bottleneck.}
The task-based view of automation~\cite{autor2003, acemoglu2018} holds that
machines substitute for labor on codifiable tasks while humans retain comparative
advantage on the rest; in our setting the residual human task is reviewing and
validating AI output, and Acemoglu and Restrepo's emphasis on labor's shifting
frontier contrasts instructively with our treatment of oversight saturation as
the binding force. The cognitive-friction term \(e^{-\lambda T}\) is empirically
motivated: randomized and field studies report large but \emph{bounded and
heterogeneous} gains from generative AI. For instance, roughly \(+15\%\) support-agent
throughput overall and \(+34\%\) for novices~\cite{brynjolfsson2025}, a
\(55.8\%\) speed-up on a programming task~\cite{peng2023}, and a \(40\%\) time
reduction with higher quality on writing tasks~\cite{noy2023}. Dell'Acqua
et al.~\cite{dellacqua2023} document a ``jagged frontier'' on which consultants
using AI \emph{off} its capability boundary were \(19\) percentage points
\emph{less} likely to be correct. That degradation-without-oversight is precisely
the mechanism our friction term encodes. We note one genuine tension: Noy and
Zhang~\cite{noy2023} find AI \emph{substitutes} for worker effort and reorients
work toward editing rather than complementing skill. Our model is consistent with
this once the two regimes are separated. Headcount and tokens are complements in
\emph{production} (more reviewers raise the volume of AI output that can be safely
shipped), yet the stochastic budget makes them \emph{substitutes at the margin}
(Section 6.2), the very behavior those experiments observe at the task level.

\paragraph{Stochastic budgeting under uncertainty.}
The probabilistic budget is a chance constraint in the sense introduced by
Charnes and Cooper~\cite{charnes1959}, and its reduction to a deterministic
equivalent follows the program they later formalized~\cite{charnes1963} and that
Pr\'ekopa~\cite{prekopa1995} develops in depth. Our approach based on a Central
Limit aggregation corrected for finite-team skewness via the Cornish--Fisher
expansion~\cite{cornish1938, jaschke2002} imports a standard tool from
value-at-risk practice, along with that tool's known limitation: the truncated
Cornish--Fisher inverse is accurate only over a bounded skewness range, so for
very heavy-tailed token usage our deterministic frontier should be read as the
tightening approximation of Assumption~2 rather than an exact boundary.

\paragraph{Human oversight and collective intelligence.}
Bainbridge's irony of automation~\cite{bainbridge1983}, adopted in
Section~1 as our central framing, is the qualitative foundation the \(e^{-\lambda T}\)
term renders quantitative. Finally, the paper speaks to
collective intelligence: where Woolley et al.~\cite{woolley2010} establish a
measurable collective-intelligence factor in human groups and Bansal et
al.~\cite{bansal2021} show that human--AI teams can beat either alone yet that
uncalibrated reliance erodes the gain, we contribute the cost side of this via the
budget-optimal \emph{composition} of a human--AI collective, and how that
composition should shift as usage volatility and skewness change.

\paragraph{Limitations: homogeneous agents, team structure, and calibration.}
The collective modeled here is \(N\) exchangeable agents: cognitive
friction depends only on per-capita intensity \(T\), which rules out
pooled review queues, dedicated reviewer roles, and
specialization---precisely the division-of-labor structures that
collective-intelligence research identifies as first-order for group
performance~\cite{woolley2010}. The empirical record likewise resists
homogeneity: novices gain roughly twice the average treatment
effect~\cite{brynjolfsson2025}, suggesting heterogeneous
\((\beta_i, \lambda_i)\). Extending the production field to a seniority
mix with a review-pooling technology, and calibrating
\((\alpha, \beta, \lambda)\) against measured usage telemetry and the
experimental effect sizes cited above, are the two most valuable next
steps. Appendix~A assembles a plausibility calibration of
\((\beta, \lambda, \sigma/\mu, \gamma_1)\) from published telemetry
and the cited effect sizes; full estimation requires panel usage data we
do not possess, so all parameters in
Figs.~\ref{fig:satiation}--\ref{fig:variance} remain illustrative.

\section*{Disclosure of Generative AI Use}

Claude (Claude Opus 4.8 and Claude Fable 5), large language models developed by
Anthropic and accessed through the Claude Code command-line interface, were used
in the preparation of this work. The author takes full responsibility for all content.

\bibliographystyle{plainnat}
\bibliography{references}

\appendix

\section{A Plausibility Calibration from Published Telemetry}

We lack the panel usage data required to \emph{estimate} the model, but
publicly reported statistics suffice to check that the illustrative
parameters of Figs.~\ref{fig:satiation}--\ref{fig:variance} occupy a
plausible region of parameter space. We record that exercise here;
nothing in the body depends on it.

\paragraph{Dispersion (\(\sigma/\mu\), \(\gamma_1\)).}
Anthropic's operator documentation for Claude Code reports two
cross-sectional cost statistics for enterprise deployments: an earlier
snapshot with a mean of \$6 per developer per active day and 90\% of
users below \$12~\cite{anthropiccosts2025}, and a later snapshot with a
mean of \$13 and 90\% below \$30~\cite{anthropiccosts2026}. Taking
cost as a proxy for token volume, each (mean, \(P_{90}\)) pair pins
down the shape of any two-parameter right-skewed family. A gamma fit
yields shape \(k = 1.78\) (\(\sigma/\mu = 0.75\),
\(\gamma_1 = 1.5\)) for the first snapshot and
\(k = 0.99\) (essentially exponential) (\(\sigma/\mu = 1.0\),
\(\gamma_1 = 2.0\)) for the second. A log-normal fit gives
\(\sigma/\mu = 0.91\), \(\gamma_1 = 3.5\) for the first snapshot
and is \emph{infeasible} for the second: no log-normal attains
\(P_{90}/\text{mean} > 2.27\), whereas the reported ratio is
\(2.31\). The published cross-sections are thus at least as dispersed
as the most dispersed compatible log-normal. Two caveats apply: these are daily
cross-sections, so monthly dispersion is lower to the extent individual
usage is serially independent (and comparable to the extent heavy users
are persistently heavy); and cost imperfectly proxies tokens under a
mixed model portfolio. Within those limits, the illustrative
\(\sigma/\mu \approx 1.1\) and \(\gamma_1 = 2\) of
Fig.~\ref{fig:tension} sit squarely inside the implied range
(\(\sigma/\mu \in [0.75, 1.0]\), \(\gamma_1 \in [1.5, 3.5]\)).

\paragraph{Output elasticity (\(\beta\)).}
Reading the cited experiments as comparisons of lower- versus
higher-intensity use, a doubling of effective AI input that raises
output by a factor \(1 + g\) implies \(\beta = \log_2(1+g)\): the
\(+15\%\) average and \(+34\%\) novice gains of Brynjolfsson et
al.~\cite{brynjolfsson2025} give \(\beta = 0.20\) and \(0.42\),
and the \(+40\%\) writing gain of Noy and Zhang~\cite{noy2023} gives
\(\beta = 0.49\). The illustrative \(\beta = 0.4\) lies in the
upper middle of this band.

\paragraph{Friction (\(\lambda\)).}
Dell'Acqua et al.~\cite{dellacqua2023} establish that the
friction-dominated regime is reachable in practice; for magnitude we
anchor the satiation intensity \(T_{\text{sat}} = \beta/\lambda\) to
the upper range of sustained productive consumption. Practitioner cost
analyses place heavy sustained usage near \(70\)~MTok per month, with
multi-agent workflows adding \(200\)--\(500\%\) token overhead
relative to the same task run single-agent~\cite{finout2026}, which is
consistent with a productive ceiling of order \(50\)--\(100\)~MTok
per engineer-month. With \(\beta = 0.4\) this brackets
\(\lambda \in [0.004, 0.008]\,\text{MTok}^{-1}\); the figures use
the upper end (\(T_{\text{sat}} = 50\)~MTok).

\paragraph{Price regime (\(P_t\)).}
The one deliberately stylized parameter is the token price. Observed
median spend of \$150--\$250 per
developer-month~\cite{anthropiccosts2026} is only \(1\)--\(2\%\) of
a loaded salary, so at today's median prices the budget trade-off is
second-order and the optimum hugs satiation. The premium regime of
Fig.~\ref{fig:tension} (\(P_t = \$300\)/MTok-equivalent) instead
describes the heavy tail in which the token line is a first-order
budget item (frontier-model multi-agent workloads, for which
four-figure single sessions and heavy-user months exceeding \$1{,}200
are already documented~\cite{finout2026}) and, prospectively,
organizations scaling per-capita compute faster than salaries. The
comparative statics of Section~6 are price-regime invariant in
direction; only the tightness of the interior optimum depends on
\(P_t\).

\end{document}